\documentclass[12pt,thmsa]{article}
\usepackage{sw20lart}

\input tcilatex
\begin{document}

\title{Comparison of UHECR spectra from necklaces and vortons\thanks{%
Presented at the Chacaltaya Meeting on Cosmic Rays Physics, La Paz, 23-27
July 2000}}
\author{Luis Masperi\thanks{%
On leave of absence from Centro At\'omico Bariloche, Argentina. E-mail:
masperi@cbpf.br} and Milva Orsaria\thanks{%
E-mail: orsaria@cbpf.br} \\
Centro Latinoamericano de F\'\i sica,\\
Av. Vencenslau Br\'az 71 Fundos, 22290-140 Rio de Janeiro, Brazil}
\date{}
\maketitle

\begin{abstract}
Cosmic rays of energy higher than $10^{19}eV$ may be explained by
topological defects produced in the early stages of universe. Two suitable
alternatives are: necklaces formed by magnetic monopoles connected by
strings, and vortons which are loops stabilized by superconducting currents.
The former are uniformly distributed in the universe, may account for cosmic
rays above the ankle, suffer a transient GZK cutoff with a subsequent
recovery and isotropy of observations is expected. The latter are
concentrated in the galactic halo, require an additional extragalactic
contribution between the ankle and the GZK cutoff, beyond which give a
harder component and predict anisotropy related to mass concentration.
\end{abstract}


\section{Introduction}

The ultra-high energy cosmic rays (UHECR) have an energy spectrum of their
flux $F(E)$ that shows for $E^3F(E)$ a minimum around $5$ x $10^{18}eV$
which is called the ankle, then a maximum before the GZK cutoff\cite
{K.Greisen} at $5$ x $10^{19}eV$ due to the interaction with the CBR and a
recovery after it at $10^{20}eV.$

Whereas the cosmic rays below the ankle are most probably of galactic
origin, it is not clear which is the explanation of the subsequent rise. The
possibility that it is due to an extragalactic source seems to be supported
by a partial GZK cutoff which would affect cosmic rays traveling at least $%
50Mpc.$ But the subsequent observed spectrum up to the highest energy event
of $3$ x $10^{20}eV$ indicates a hard component which may be or not related
to that which appears above the ankle.

It is difficult to explain the observed events\cite{M.Takeda} beyond the GZK
cutoff with ordinary astrophysical objects which are not identified if they
are close to us, and that would require non standard messengers not
interacting with CBR if they are very far away\cite{A.V.Olinto} .

A solution may be the top-down mechanism where some type of superheavy
microscopical object with mass of the order of the Grand Unification Theory
(GUT) scale decays very slowly producing the observed UHECR. Since a general
feature is a hard spectrum at emission $F(E_{em})\propto 1/E_{em}$ , these
superheavy objects might be either condensed in the galactic halo or
uniformly distributed in the universe.

The former case may correspond to closed cosmic strings stabilized by
superconducting currents called vortons\cite{R.L.Davis} or to superheavy
particles whose interaction with the ordinary ones is of gravitational order
generically denoted as cryptons\cite{V.A.Kuzmin} . Since they were
presumably produced at the GUT scale, they behave afterwards as cold dark
matter (CDM) and should have concentrated in the galactic halo.

The latter case may be instead represented by necklaces\cite{V.Berezinski}
where ordinary cosmic strings whose dynamics makes them evolve to a scaling
solution, i.e. a uniform distribution in space, incorporate monopoles and
antimonopoles that annihilate very slowly.

We will give a brief description of vortons and particularly show that if
they represent a small fraction of the halo CDM, they may account for the
apparent hard component of the UHECR spectrum above $10^{20}eV.$ An
additional extragalactic component would be necessary to explain the ankle
feature.

Compared to this, an also small fraction of the critical density of universe
represented by necklaces may give, with a reasonable law for the energy
degrading due to interaction with CBR, the maximum of $E^3F(E)$ immediately
below the GZK cutoff. The hard spectrum at emission allows a recovery above
it at least up to the highest observed UHECR which would be impossible for
extragalactic sources with the ordinary law $F(E_{em})\propto 1/E_{em}^3$ .

\section{Vortons in halo}

Considering sources that emit $\ \stackrel{.}{n}_X\left( t\right) \quad $ $%
GUT$ boson particles $X$ per unit space and time, each of them giving $N_c$
UHECR, the total flux on earth will be 
\begin{equation}
F=\frac 1{4\pi }\int_{t_{in}}^{t_0}dt\text{ }N_c\stackrel{.}{n}_X(t)\left( 
\frac{a(t)}{a(t_0)}\right) ^3\text{ ,}  \label{e1}
\end{equation}
where $a$ is the scale parameter of universe, $t_0$ its age and $t_{in}$ the
initial time of contributions.

For quasistable objects like vortons concentrated in halo with density $n(t)$
which emit by tunneling an $X$ with a lifetime $\tau ,\quad \stackrel{.}{n}%
_X=n/\tau $ and the total flux is 
\begin{equation}
F_h=\frac{N_c}{4\pi }n_h(t_0)\frac{\Delta t}\tau \text{ ,}  \label{e2}
\end{equation}
where $\Delta t\sim 50kpc$ is the halo size.

The energy spectrum 
\begin{equation}
F_h(E)=\frac 1{4\pi }n_h(t_0)\frac{\Delta t}\tau \sum_{i=1}^{N_c}\delta
(E-E_i)  \label{e3}
\end{equation}
must be averaged on the intervals $\Delta E_i$ between produced particles to
compare with observations 
\begin{equation}
\overline{F}_h(E_i)=\frac 1{\Delta E_i}\text{ }\frac{n_h(t_0)}{4\pi }\text{ }%
\frac{\Delta t}\tau \text{ .}  \label{e4}
\end{equation}

Since the production of UHECR comes from the hadronization of the very
energetic quark into which the $X$ particle decays, by dimensional arguments
there are no relevant ordinary mass parameters and we may expect 
\begin{equation}
\Delta E_i\sim E_i\text{ , }\overline{F}_h(E_i)\propto \frac 1{E_i}\text{ ,}
\label{e5}
\end{equation}
i.e. a hard spectrum consistent with accurate QCD calculations\cite{M.Birkel}
apart from energies close to $m_X.$

It is interesting that according to Eq.(5) the average probability of UHECR
production on energy intervals will be 
\begin{equation}
\frac{d\Gamma }{dE}\sim \frac 1\tau \text{ }\frac 1E\text{ ,}  \label{e6}
\end{equation}
which integrated on the ultra-high energy range $10^{19}-10^{24}eV$ to agree
with the total flux Eq.(2) must give 
\begin{equation}
\Gamma =\int \frac{d\Gamma }{dE}dE=\frac{N_c}\tau \text{ ,}  \label{e7}
\end{equation}
with $N_c\sim 10$ , that is a reasonably accepted value\cite{L.Masperi} .

Therefore we may take the equally spaced particles in $\log E$ according to

$E_1\simeq 10^{19}eV,$ $E_2\simeq 10^{19.5}eV,$ $E_3\simeq 10^{20}eV,$ $%
E_4\simeq 10^{20.5}eV.............E_{10}\simeq 10^{23.5}eV$

so that $\Delta E_i\simeq 0.7E_i\quad .$

Since from eq.(4) the flux in each energy bin is the same, one may normalize
it roughly at the expected value for $10^{20}eV$%
\begin{equation}
\frac{n_h(t_0)}{4\pi }\text{ }\frac{\Delta t}\tau =\frac 1{km^2\text{ }%
century}\text{ .}  \label{e8}
\end{equation}

A vorton is a loop of ordinary cosmic string with an energy per unit length $%
\mu \sim m_X^2$ stabilized by $N$ massless fermionic carriers giving
therefore a total energy 
\begin{equation}
E_v=\mu L+\frac{N^2}L\text{ ,}  \label{e9}
\end{equation}
which is minimized at $E_v\sim 2Nm_X$ by a length $L\sim N/m_X$ . The decay
of the vorton with emission of $X$ by tunneling gives a lifetime $\tau \sim
t_0$ for $N\sim 1000.$ Therefore to satisfy Eq.(8) one only needs a fraction 
$\sim 10^{-6}$ of the average energy density of the halo $0.3$ $GeV/cm^3$
represented by vortons. This small density may be the remnant of the
collapse of most vortons at the electroweak phase transition\cite{L.Masperi}
.

This contribution of vortons to $E^3F(E)$ will be a hard component which
reproduces the observed flux above the GZK cutoff as is seen in fig.1. One
must therefore complement it with another possibly extragalactic component
which may explain the spectrum between the ankle and GZK energy.

A similar analysis may be done for cryptons being necessary to explain their
required density.

\section{Necklaces in universe}

These hybrid topological defects may be formed by a sequence of GUT symmetry
breakings 
\begin{equation}
G\rightarrow H\text{ x }U(1)\rightarrow H\text{ x }Z_2\text{ ,}  \label{e10}
\end{equation}
where in the first monopoles would be produced and then would be attached to
the ordinary strings which appear in the second one as beads of a necklace.

The relevant parameter for the necklace dynamics is $r=m/(\mu d)$ where $m$
is the monopole mass, $d$ its separation from the antimonopole in the string
and $\mu $ the tension of the latter. For $r\sim 10^6$ the distance between
strings at present is small $\sim 3Mpc.$

The evolution of the necklace networks is scale invariant, $i.e.$ they would
be distributed uniformly in the universe and represent a constant fraction
of its energy. Monopoles and antimonopoles trapped in the necklaces at the
end would annihilate producing $X$ particles with a rate 
\begin{equation}
\stackrel{.}{n}_X\left( t\right) \sim \frac{r^2\text{ }\mu }{t^3\text{ }m_X}%
=\frac \alpha {t^3}\cdot  \label{e11}
\end{equation}

Therefore the expression for the UHECR flux eq.(1) would apply but with an
early $t_{in}$ compatible with avoiding their redshift below $10^{19}eV$
which is roughly of the order of the matter-radiation equivalence time. $%
r^2\mu $ cannot be larger than $10^{28}GeV^2$ to avoid a diffuse gamma
radiation above the experimental bound. The total flux will be 
\begin{equation}
F_u=N_c\ \frac \alpha {t_0^2}\ \ln \left( \frac{t_0}{t_{in}}\right) \cdot
\label{e12}
\end{equation}

Even though at emission the UHECR produced by an $X$ are equally spaced in $%
\log E_{em}$ as discussed in Sec.2, their redshift would cause a softening
of the law $1/E$ for the flux spectrum on earth.

But more important than this effect is the interaction of cosmic rays with
CBR. The $p\gamma $ total cross-section at the highest energy is $\sim 0.2$ $%
mb$ , and rises up to $\sim 0.6$ $mb$ for the $\Delta $ resonance mass.

Then to evaluate the flux spectrum, we may proceed as follows. Instead of
taking one cosmic ray in each bin as in the case of vortons in halo, we will
consider $N_i$ to account for the degrading of energy so that

\begin{equation}
\overline{F}_U\left( E_i\right) =\frac \alpha {t_0^2}\ \ln \left( \frac{t_0}{%
t_{in}}\right) \frac{N_i}{\Delta E_i}\cdot  \label{e13}
\end{equation}

Therefore we may parameterize for all cases 
\begin{equation}
\log \left[ E_i^3\overline{F}(E_i)\right] =\log J+\log N_i+2\log \left( 
\frac{E_i}{10^{19}eV}\right) \text{ },  \label{e14}
\end{equation}
where $J$ , related to the properties of sources, will be adjusted to fit
the observed events. For the case of vortons all $N_i=1.$

For necklaces to determine the effective $N_i$ and considering the mean free
path associated to the quoted $\sigma _{p\gamma }$ and the density of $3K$
photons, we will take $1\%$ of probability that the cosmic ray for $\Delta $
production keeps its energy. This is consistent with the fact that the
sources of non degrading protons are concentrated in a radius $\sim 50Mpc$
out a whole space $\sim 100$ times larger. For higher $E_i$ this probability
will increase to $3\%$ following $\sigma _{p\gamma }.$ We will assume that
the missing events are transferred in equal parts to the two inmediate lower
bins. According to the bins defined in Sec.2, that of $E_3$ is particularly
affected by the resonant scattering. The inmediate lower one has only the
upper $10\%$ of the bin in the resonance region so that $\sim 90\%$ of its
events keep their energy and the rest is transferred to the bin of $E_1$ .

In this way it turns out

$N_1=4.09$ $N_2=5.15$ $N_3=0.05$ $N_4=0.14$ $N_5=0.12$

$N_6=0.10$ $N_7=0.08$ $N_8=0.06$ $N_9=0.04$ $N_{10}=0.03$

and the flux spectrum eq.(14) is shown in fig.2.

We see that the existence of an ankle and the recovery after a transient GZK
are successfully reproduced.

A check of the calculation is that the normalization for the bin of $%
10^{19}eV$ must be 
\begin{equation}
\frac \alpha {t_0^2}\ \ln \left( \frac{t_0}{t_{in}}\right) \ N_1=\frac 1{km^2%
\text{ }yr}\quad ,  \label{e15}
\end{equation}
which is satisfied for $\alpha =10^{37}sec^{-1}$ coming from $%
m_X=10^{15}GeV. $

This is similar to the normalization for ordinary strings\cite
{P.Bhattacharjee} the difference being that for them $\mu \sim m_X^2$ and
the present separation between strings is three orders of magnitude larger
than for necklaces. As a consequence our simplified treatment is much more
suitable for the latter because protons for sources within a radius $\sim
50Mpc$ would be detected whereas photons would be mostly absorbed.

It is clear that the feature of fig.2 is consequence of the hard component
corresponding to quark hadronization reflected in the last term of eq.(14).
If one should have considered an ordinary law $1/E^3$ at emission, the
corresponding flux due to uniformly distributed extragalactic sources would
be given by eq.(14) without the last term and as shown in fig.3 would
reasonably reproduce the observed events below the GZK cutoff but without
recovery above it.

We must note comparing figs. 1 and 2 that $J$ is one order of magnitude
larger for necklaces than for vortons which is consequence of the fact that
the latter fit the flux at $\sim 10^{20}eV$ and the former that at $\sim
10^{19}eV$ with a partial compensation due to degrading of energy. It is
important to note that with the above values of $r,$ $\mu $ and a monopole
mass $m\sim 10^{16}GeV$ the energy per unit length due to monopoles turns
out to be $\sim 10^{22}GeV^2$ and the fraction of critical density $\sim
10^{-9},$ slightly smaller than that of ordinary strings.

\section{Conclusions}

We have seen that both vortons in halo and necklaces in universe may be a
solution for the problem of UHECR above $10^{20}eV.$ From the observational
point of view the difference will be the expected anisotropy in the former
case because of the asymmetric position of the sun in the galaxy compared to
the isotropy of the latter characteristic of a cosmological origin.
Regarding this, it must be noted that the anisotropy detected below the
ankle is not observed above it\cite{M.Ta} , which would be consistent with
the appearance of an extragalactic component, a larger statistics being
needed at the highest energy to see if new galactic sources contribute.

Referring to elementary particle theory it is interesting that the most
appropriate GUT models are different. For necklaces, since it is necesary
that the breaking of the abelian symmetry leaves a discrete $Z_2$ unbroken,
a GUT model based on $SO(10)$ is suitable. For vortons on the other hand the 
$E_6$ GUT model is better because the breaking of the contained additional
abelian symmetry produces the necessary superconducting current with exotic
fermions, whereas necklaces would not be formed due to the fact that its
Higgs content does not allow an unbroken $Z_2$ .

\begin{center}
\textbf{Acknowledgement}
\end{center}

We are grateful to Pedro Miranda and the organizing committee of the
Chacaltaya Meeting on Cosmic Rays Physics for the warm hospitality at La
Paz. MO thanks Fundaci\'on Antorchas for partial financial support.

\end{document}